# A measure of association between vectors based on "similarity covariance"

RD Pascual-Marqui[1,2], D Lehmann[2], K Kochi[2], T Kinoshita[3], N Yamada[1]

[1]Department of Psychiatry, Shiga University of Medical Science, Japan; [2]The KEY Institute for Brain-Mind Research, University of Zurich, Switzerland; [3]Department of Neuropsychiatry, Kansai Medical University, Japan.

Corresponding author:
RD Pascual-Marqui; pascualm *at* key.uzh.ch; pascualm *at* belle.shiga-med.ac.jp; www.uzh.ch/keyinst/loreta.htm
The KEY Institute for Brain-Mind Research, University Hospital of Psychiatry, Zurich, Switzerland
Department of Psychiatry, Shiga University of Medical Sciences, Shiga, Japan

## 1. Abstract

The "maximum similarity correlation" definition introduced in this study is motivated by the seminal work of Székely et al on "distance covariance" (Ann. Statist. 2007, 35: 2769-2794; Ann. Appl. Stat. 2009, 3: 1236-1265). Instead of using Euclidean distances "$d$" as in Székely et al, we use "similarity", which can be defined as "exp(-d/s)", where the scaling parameter s>0 controls how rapidly the similarity falls off with distance. Scale parameters are chosen by maximizing the similarity correlation. The motivation for using "similarity" originates in spectral clustering theory (see e.g. Ng et al 2001, Advances in Neural Information Processing Systems 14: 849-856). We show that a particular form of similarity correlation is asymptotically equivalent to distance correlation for large values of the scale parameter. Furthermore, we extend similarity correlation to coherence between complex valued vectors, including its partitioning into real and imaginary contributions. Several toy examples are used for comparing distance and similarity correlations. For instance, points on a noiseless straight line give distance and similarity correlation values equal to 1; but points on a noiseless circle produces near zero distance correlation (dCorr=0.02) while the similarity correlation is distinctly non zero (sCorr=0.36). In distinction to the distance approach, similarity gives more importance to small distances, which emphasizes the local properties of functional relations. This paper represents a preliminary empirical study, showing that the novel similarity association has some distinct practical advantages over distance based association.

For the sake of reproducible research, the software code implementing all methods discussed here (using lazarus free-pascal "www.lazarus.freepascal.org"), including the test data as text files are freely available at:
sites.google.com/site/pascualmarqui/home/similaritycovariance

# Contents

## 2. The sample distance covariance and correlation of Székely et al ([1] and [2])

Consider the *r*-dimensional multivariate vector $\mathbf{Z}_i \in \mathbb{R}^{r \times 1}$, for which there are *N* independent samples, i.e. $i = 1...N$.

Next, consider a partition of the variables contained in $\mathbf{Z}_i$, into two groups, corresponding to $\mathbf{X}_i \in \mathbb{R}^{p \times 1}$ and $\mathbf{Y}_i \in \mathbb{R}^{q \times 1}$ with $(r = p + q)$.

Next, consider the symmetric matrices $\mathbf{a} \in \mathbb{R}^{N \times N}$ and $\mathbf{b} \in \mathbb{R}^{N \times N}$ defined as:

**Eq. 1** $\quad a_{ij} = |\mathbf{X}_i - \mathbf{X}_j|$

**Eq. 2** $\quad b_{ij} = |\mathbf{Y}_i - \mathbf{Y}_j|$

for $(i, j) = 1...N$, corresponding to the Euclidean distances between pairs of samples.

Next, consider the double-centered matrices:

**Eq. 3** $\quad \mathbf{A} = \mathbf{HaH} \quad \in \mathbb{R}^{N \times N}$

**Eq. 4** $\quad \mathbf{B} = \mathbf{HbH} \quad \in \mathbb{R}^{N \times N}$

where:

**Eq. 5** $\quad \mathbf{H} = \left( \mathbf{I} - \frac{\mathbf{1}\mathbf{1}^T}{N} \right) \in \mathbb{R}^{N \times N}$

is the centering matrix, $\mathbf{I} \in \mathbb{R}^{N \times N}$ is the identity matrix, $\mathbf{1} \in \mathbb{R}^{N \times 1}$ is a vector of ones, and the superscript "*T*" denotes transposed. Note that for a double centered matrix, the sum of elements of any column is zero, and the sum of elements of any row is zero.

The distance covariance between the *p* variables in **X** and the *q* variables in **Y** is defined as:

**Eq. 6** $\quad \mathcal{V}_d(\mathbf{X}, \mathbf{Y}) = \frac{1}{N^2} tr(\mathbf{AB}) = \frac{1}{N^2} \sum_{i=1}^{N} \sum_{j=1}^{N} A_{ij} B_{ij}$

where "$tr(\cdot)$" denotes the trace of the argument.
The distance variance of **X** is:

**Eq. 7** $\quad \mathcal{V}_d(\mathbf{X}, \mathbf{X}) = \frac{1}{N^2} tr(\mathbf{A}^2) = \frac{1}{N^2} \sum_{i=1}^{N} \sum_{j=1}^{N} A_{ij}^2$

and similarly, the distance variance of **Y** is:

**Eq. 8** $\quad \mathcal{V}_d(\mathbf{Y}, \mathbf{Y}) = \frac{1}{N^2} tr(\mathbf{B}^2) = \frac{1}{N^2} \sum_{i=1}^{N} \sum_{j=1}^{N} B_{ij}^2$

The distance correlation between the two sets of variables is defined as:

**Eq. 9** $\quad \mathcal{R}_d(\mathbf{X}, \mathbf{Y}) = \begin{cases} \dfrac{\mathcal{V}_d(\mathbf{X}, \mathbf{Y})}{\sqrt{\mathcal{V}_d(\mathbf{X}, \mathbf{X}) \mathcal{V}_d(\mathbf{Y}, \mathbf{Y})}}, & \text{if } \mathcal{V}_d(\mathbf{X}, \mathbf{X}) \mathcal{V}_d(\mathbf{Y}, \mathbf{Y}) > 0 \\ 0, & \text{if } \mathcal{V}_d(\mathbf{X}, \mathbf{X}) \mathcal{V}_d(\mathbf{Y}, \mathbf{Y}) = 0 \end{cases}$

which equivalently is:

**Eq. 10**
$$\mathcal{R}_d(\mathbf{X},\mathbf{Y}) = \begin{cases} \dfrac{tr(\mathbf{AB})}{\sqrt{tr(\mathbf{A}^2)tr(\mathbf{B}^2)}}, & \text{if } tr(\mathbf{A}^2)tr(\mathbf{B}^2) > 0 \\ 0, & \text{if } tr(\mathbf{A}^2)tr(\mathbf{B}^2) = 0 \end{cases}$$

These definitions (Eq. 1 to Eq. 10) correspond to the sample versions of "distance" covariance (Eq. 6), variance (Eq. 7 and Eq. 8), and correlation (Eq. 9 and Eq. 10) of Székely et al ([1] and [2]), denoted with a subscript "d" as $\mathcal{V}_d$ and $\mathcal{R}_d$.

This gives a generalized covariance between the two sets of variables, which is not limited to linear relations, and which takes the value zero only under complete statistical independence ([1] and [2]).

## 3. The new maximum similarity correlation

Consider the symmetric similarity matrices $\mathbf{d} \in \mathbb{R}^{N \times N}$ and $\mathbf{e} \in \mathbb{R}^{N \times N}$, with elements:

**Eq. 11** $\quad d_{ij} = \exp\left(-|\mathbf{X}_i - \mathbf{X}_j|^\alpha / s_x\right) = \exp\left(-a_{ji}^\alpha / s_x\right)$

**Eq. 12** $\quad e_{ij} = \exp\left(-|\mathbf{Y}_i - \mathbf{Y}_j|^\alpha / s_y\right) = \exp\left(-b_{ji}^\alpha / s_x\right)$

where $\alpha > 0$, $s_x > 0$, $s_y > 0$ are parameters; and where $a_{ij}$ and $b_{ij}$ are the Euclidean distances defined in Eq. 1 and Eq. 2.

Similarity matrices are used, e.g. in the field of spectral clustering (see e.g. Ng et al [3]).

Next, consider the triple-centered matrices of dimension $\mathbb{R}^{N \times N}$ defined as:

**Eq. 13** $\quad \mathbf{D} = \mathbf{HdH} - \left[\dfrac{tr(\mathbf{HdH})}{N-1}\right]\mathbf{H}$

**Eq. 14** $\quad \mathbf{E} = \mathbf{HeH} - \left[\dfrac{tr(\mathbf{HeH})}{N-1}\right]\mathbf{H}$

where $\mathbf{H}$ is the centering matrix (Eq. 5).

This form of double-centering (Eq. 13 and Eq. 14) is more general than the one used by Székely et al ([1] and [2]) in Eq. 3 and Eq. 4. The matrices $\mathbf{D}$ and $\mathbf{E}$ retain the main property of having each row and each column with zero means, but at the same time, they minimize the similarity variances, as will be shown later (see Eq. 20 to Eq. 25). The additional term in the left hand side of Eq. 13 and Eq. 14 might reduce bias (see e.g. [7]).

The similarity variance of $\mathbf{X}$ is:

**Eq. 15** $\quad \mathcal{V}_s(\mathbf{X},\mathbf{X};\alpha,s_x) = \dfrac{1}{N^2} tr(\mathbf{D}^2) = \dfrac{1}{N^2} \sum_{i=1}^{N}\sum_{j=1}^{N} D_{ij}^2$

and similarly, the similarity variance of $\mathbf{Y}$ is:

**Eq. 16** $$\mathcal{V}_s(\mathbf{Y},\mathbf{Y};\alpha,s_y) = \frac{1}{N^2} tr(\mathbf{E}^2) = \frac{1}{N^2}\sum_{i=1}^{N}\sum_{j=1}^{N} E_{ij}^2$$

The similarity covariance between the $p$ variables in $\mathbf{X}$ and the $q$ variables in $\mathbf{Y}$ is defined as:

**Eq. 17** $$\mathcal{V}_s(\mathbf{X},\mathbf{Y};\alpha,s_x,s_y) = \frac{1}{N^2} tr(\mathbf{DE}) = \frac{1}{N^2}\sum_{i=1}^{N}\sum_{j=1}^{N} D_{ij} E_{ij}$$

This notation emphasizes the dependence on the parameters $\alpha > 0$, $s_x > 0$, and $s_y > 0$.

For a given exponent parameter $\alpha > 0$, the similarity correlation between the two sets of variables is defined as:

**Eq. 18** $$\mathcal{R}_s(\mathbf{X},\mathbf{Y}) = \begin{cases} \max_{s_x,s_y} \dfrac{\mathcal{V}_s(\mathbf{X},\mathbf{Y};\alpha,s_x,s_y)}{\sqrt{\mathcal{V}_s(\mathbf{X},\mathbf{X};\alpha,s_x)\mathcal{V}_s(\mathbf{Y},\mathbf{Y};\alpha,s_y)}}, & \text{if } \mathcal{V}_s(\mathbf{X},\mathbf{X};\alpha,s_x)\mathcal{V}_s(\mathbf{Y},\mathbf{Y};\alpha,s_y) > 0 \\ 0, & \text{if } \mathcal{V}_s(\mathbf{X},\mathbf{X};\alpha,s_x)\mathcal{V}_s(\mathbf{Y},\mathbf{Y};\alpha,s_y) = 0 \end{cases}$$

In words: this means that the similarity correlation is the maximum correlation based on the similarity covariance and variances, with respect to the scale parameters $(s_x,s_y)$.

These definitions (Eq. 11 to Eq. 18) correspond to the sample versions of "similarity" covariance (Eq. 17), variance (Eq. 15 and Eq. 16), and correlation (Eq. 18), denoted with a subscript "$s$" as $\mathcal{V}_s$ and $\mathcal{R}_s$.

## 4. Notes on the maximum similarity correlation

### a. On the exponent $\alpha$ for the Euclidean distance

Typical values for the exponent $\alpha > 0$ in Eq. 11 and Eq. 12 can be $\alpha = 1$ and $\alpha = 2$. The effect of using different $\alpha$ values has not been studied at the time of this writing.

### b. Alternative kernels

In Eq. 11 and Eq. 12, other kernel choices are possible such as:

**Eq. 19** $$d_{ij} = \begin{cases} \left(1 - |\mathbf{X}_i - \mathbf{X}_j|^\alpha / s_x\right)^2, & \text{if } \left(|\mathbf{X}_i - \mathbf{X}_j|^\alpha / s_x\right) < 1 \\ 0, & \text{otherwise} \end{cases}$$

A similar definition is used for the elements of $\mathbf{e}$ based on the data $\mathbf{Y}$.

For the case $\alpha = 2$, this kernel (Eq. 19) is equal in form to Tukey's biweight (see e.g. Maronna et al [4]).

### c. On the particular choice of triple-centering of the similarity matrices

The triple-centering definition used in Eq. 13 and Eq. 14 is more general than the one used by Székely et al ([1] and [2]) corresponding to Eq. 3 and Eq. 4. The matrices **D** and **E** retain the main property of having each row and each column with zero means, but at the same time, they minimize the similarity variances. The derivation follows.

Consider, for instance, the similarity matrix **d** for the data **X**, and the more general form of double centering:

Eq. 20 $\quad \mathbf{D} = \mathbf{HdH} - \beta \mathbf{H}$

where the parameter $\beta$ is an arbitrary scalar. The matrix **D** satisfies the conditions of double centering, i.e:

Eq. 21 $\quad \begin{cases} \mathbf{1}^T \mathbf{D} = \mathbf{0}^T \\ \mathbf{D1} = \mathbf{0} \end{cases}$

The parameter $\beta$ in Eq. 20 can be interpreted as a scalar mean value (multiplying **H**), which should optimally be chosen to minimize the variance:

Eq. 22 $\quad \mathcal{V}_s(\mathbf{X},\mathbf{X};\alpha,s_x) = \min_{\beta} \frac{1}{N^2} tr(\mathbf{D}^2) = \min_{\beta} \frac{1}{N^2} tr\left([\mathbf{HdH} - \beta \mathbf{H}]^2\right)$

which has solution:

Eq. 23 $\quad \beta = \frac{tr(\mathbf{HdH})}{tr(\mathbf{H})} = \frac{tr(\mathbf{HdH})}{N-1}$

The additional term in the left hand side of Eq. 20 might reduce bias (see e.g. [7]).

This result guarantees that the similarity variance is zero in the two extreme degenerate cases when the scale parameter $s_x \to 0$ or $s_x \to \infty$. Note that:

Eq. 24 $\quad \begin{cases} \lim_{s_x \to 0} \mathbf{d} = \mathbf{I} \\ \lim_{s_x \to \infty} \mathbf{d} = \mathbf{11}^T \end{cases}$

which produce:

Eq. 25 $\quad \begin{cases} \lim_{s_x \to 0} \left\{ \mathbf{HdH} - \left[\frac{tr(\mathbf{HdH})}{N-1}\right]\mathbf{H} \right\} = \mathbf{H} - \mathbf{H} = \mathbf{0} \\ \lim_{s_x \to \infty} \left\{ \mathbf{HdH} - \left[\frac{tr(\mathbf{HdH})}{N-1}\right]\mathbf{H} \right\} = \mathbf{0} - \mathbf{0} = \mathbf{0} \end{cases}$

Similar results hold for the data **Y**.

### d. On the asymptotic equivalence between distance correlation and a modified similarity correlation

In Eq. 11 and Eq. 12, consider the case when $\alpha = 1$, and when $s_x$ and $s_y$ are both sufficiently large. Then the following approximations hold:

**Eq. 26**
$$\begin{cases} d_{ij} = \exp\left(-|\mathbf{X}_i - \mathbf{X}_j|/s_x\right) \approx 1 - |\mathbf{X}_i - \mathbf{X}_j|/s_x \\ e_{ij} = \exp\left(-|\mathbf{Y}_i - \mathbf{Y}_j|/s_y\right) \approx 1 - |\mathbf{Y}_i - \mathbf{Y}_j|/s_y \end{cases}$$

Thus:

**Eq. 27**
$$\begin{cases} \mathbf{d} = \mathbf{1}\mathbf{1}^T - \dfrac{1}{s_x}\mathbf{a} \\ \mathbf{e} = \mathbf{1}\mathbf{1}^T - \dfrac{1}{s_y}\mathbf{b} \end{cases}$$

where $\mathbf{a}$ and $\mathbf{b}$ are the distance matrices (Eq. 1 and Eq. 2). If use is made of the classical double-centering as defined in Székely et al ([1] and [2]) (see Eq. 3 and Eq. 4), then Eq. 27 gives the following modified similarity matrices:

**Eq. 28**
$$\begin{cases} \mathbf{A}' = \mathbf{H}\mathbf{d}\mathbf{H} = -\dfrac{1}{s_x}\mathbf{A} \\ \mathbf{B}' = \mathbf{H}\mathbf{e}\mathbf{H} = -\dfrac{1}{s_y}\mathbf{B} \end{cases}$$

By using these modified double centered similarity matrices (Eq. 28) in Eq. 6-Eq. 9, it can now be easily shown that the modified similarity correlation is asymptotically equal to the distance correlation. This result shows the close connection between the distance and similarity covariances.

### e. On maximizing similarity covariance and similarity variance with respect to the scale parameters

The definition of similarity correlation in Eq. 18 includes an explicit maximization with respect to the scale parameters $(s_x, s_y)$.

However, in an independent manner, and for whatever other purposes, one can independently calculate the maxima with respect to the scale parameters of:
1. The similarity covariance
2. The similarity variances for **X** and **Y**.

For instance, consider the problem of finding clusters in a single data set such as **X**. And consider the method of spectral clustering, which is based on the similarity matrix of **X** (actually the method uses some form of the Laplacian matrix, derived from the similarity matrix [3]). In spectral clustering publications very little attention is given to the choice of the scale parameter. We speculate that a good choice for the scale parameter is the one that maximizes the similarity variance $\mathcal{V}_s(\mathbf{X},\mathbf{X};\alpha,s_x)$ of Eq. 15, based on Eq. 11 and Eq. 13.

### f. On related previous work and distinctions [14]

Note that the similarity covariance defined in Eq. 17 has very similar form to the "Hilbert-Schmidt independence criterion" (HSIC) statistic defined by Gretton et al in 2005 [11].

However, there are a number of differences that make our work, with its aims, results, and novelty, very distinct from the work of Gretton et al:
1. In our work, the only role played by similarity covariance is its use for defining the maximum similarity correlation (and coherence), which are our statistics of interest. No other use is made of similarity covariance in this work.
2. We use a triple centering form (Eq. 13 and Eq. 14) which is **_distinct_** from the double centering form implicit in HSIC.
3. HSIC [11] uses fixed values for the scale parameters. In sharp contrast, we do not make any direct use of the covariance (Eq. 17), but rather of the correlation (Eq. 18), which is maximized with respect to the scale parameters.
4. HSIC with Gaussian ($\alpha$=2) and Laplace ($\alpha$=1) kernels always degenerates with very large or very small scale values (Eq. 24), while the maximum similarity correlation does not.

It is only fair to emphasize that the similarity correlation proposed here has a similar but distinct structure to the kernel target alignment of Cristianini et al (2001) [12], but see also Cortes et al (2012) [13]. The distinctions consist of the different triple-centering used here (Eq. 13 and Eq. 14), and the maximization of the similarity correlation with respect to the scale parameter, which has not been reported previously, to the best of our knowledge.

It is also fair to emphasize that the underlying structure for the similarity correlation has been long-time published by Escoufier (1973) [15], as the RV coefficient:

**Eq. 29** $$RV(\mathbf{X},\mathbf{Y}) = \frac{tr\left[(\mathbf{H}\mathbf{X}^T\mathbf{X}\mathbf{H})(\mathbf{H}\mathbf{Y}^T\mathbf{Y}\mathbf{H})\right]}{\sqrt{tr\left[(\mathbf{H}\mathbf{X}^T\mathbf{X}\mathbf{H})^2\right] tr\left[(\mathbf{H}\mathbf{Y}^T\mathbf{Y}\mathbf{H})^2\right]}}$$

where $\mathbf{X} \in \mathbb{R}^{p \times N}$ and $\mathbf{Y} \in \mathbb{R}^{q \times N}$ are the data matrices whose columns consist of the $N$ samples. Note that $(\mathbf{X}^T\mathbf{X})$ consists of similarities between all pairs of samples expressed as the scalar product $(\mathbf{X}_i^T\mathbf{X}_j)$. In distinction, the similarity correlation:

1. uses in place of $(\mathbf{X}^T\mathbf{X})$ another form of similarity that decreases with the distance between the pairs of samples (Eq. 11 and Eq. 12);
2. uses triple-centering of the similarity matrices, not double-centering;
3. is maximized with respect to the scale parameters;
4. is not limited to linear relations, as is the case of the RV coefficient.

## 5. A degeneracy condition for the distance and similarity correlations

Consider the case where all samples for the data set in **X** are equidistant from each other. An example consists of four equidistant points in $\mathbb{R}^3$, on the vertices of a regular tetrahedron. In such a case, for any arbitrary data set **Y**, all permutations of the samples of **X** will produce the same distance covariance and similarity covariance values. Unfortunately, this is a meaningless situation, unless the covariance value is zero. It will now be shown that while the distance covariance is non-zero (which can be misleading), the similarity covariance is zero.

In this degenerate case, the distance matrix has the following structure:

**Eq. 30** $\quad \mathbf{a} = a\left(\mathbf{1}\mathbf{1}^T - \mathbf{I}\right)$

where $a$ is the distance between any two points. The double centered distance matrix is:

**Eq. 31** $\quad \mathbf{A} = \mathbf{H}\mathbf{a}\mathbf{H} = a\mathbf{H}\left(\mathbf{1}\mathbf{1}^T - \mathbf{I}\right)\mathbf{H} = -a\mathbf{H}$

This means that the distance covariance for any data set **Y** will be non-zero, which can be misleading.

In this same degenerate case, now consider similarity matrix, which has the following structure:

**Eq. 32** $\quad \mathbf{d} = \mathbf{I} + d\left(\mathbf{1}\mathbf{1}^T - \mathbf{I}\right)$

with $d = \exp(-a/s_x)$. Note that this matrix has unit diagonal elements, and all off-diagonal elements equal to $d$. The new triple centered (Eq. 13) similarity matrix is zero:

**Eq. 33** $\quad \mathbf{D} = \mathbf{H}\mathbf{d}\mathbf{H} - \left[\dfrac{tr(\mathbf{H}\mathbf{d}\mathbf{H})}{N-1}\right]\mathbf{H} = (1-d)\mathbf{H} - (1-d)\mathbf{H} = \mathbf{0}$

Thus, the similarity variance of **X** is zero, which will then produce zero values for the covariance and correlation, which at least is not misleading.

## 6. Complex valued data: the new maximum similarity coherence

Covariances for complex valued random variables are common in time series analysis, see e.g. Brillinger 1981 [5].

The basic definition for similarity covariance can be extended to complex valued data, now with $\mathbf{X}_i \in \mathbb{C}^{p \times 1}$ and $\mathbf{Y}_i \in \mathbb{C}^{q \times 1}$, for $i = 1...N$ independent samples. The similarities are defined as in Eq. 11 and Eq. 12, now as:

**Eq. 34** $\quad \begin{cases} d_{ij} = \exp\left(-a_{ji}^{\alpha}/s_x\right) \\ e_{ij} = \exp\left(-b_{ji}^{\alpha}/s_x\right) \end{cases}$

with:

**Eq. 35**
$$\begin{cases} a_{ij} = |\mathbf{X}_i - \mathbf{X}_j| = \sqrt{(\mathbf{X}_i - \mathbf{X}_j)^*(\mathbf{X}_i - \mathbf{X}_j)} = \sqrt{|\text{Re}\mathbf{X}_i - \text{Re}\mathbf{X}_j|^2 + |\text{Im}\mathbf{X}_i - \text{Im}\mathbf{X}_j|^2} \\ b_{ij} = |\mathbf{Y}_i - \mathbf{Y}_j| = \sqrt{(\mathbf{Y}_i - \mathbf{Y}_j)^*(\mathbf{Y}_i - \mathbf{Y}_j)} = \sqrt{|\text{Re}\mathbf{Y}_i - \text{Re}\mathbf{Y}_j|^2 + |\text{Im}\mathbf{Y}_i - \text{Im}\mathbf{Y}_j|^2} \end{cases}$$

where the superscript "*" denotes transposed and complex conjugated, and where "Re" and "Im" denote real and imaginary parts respectively. Note that the real and imaginary vectors in the extreme right hand side of Eq. 35 are real-valued vectors.

The similarity covariance (Eq. 17) calculated with these definitions (Eq. 34 and Eq. 35), denoted as $\mathcal{V}_s(\mathbf{X}, \mathbf{Y})$, corresponds to the total similarity covariance for complex valued data, which includes all relations between real and imaginary parts.

The total similarity coherence, denoted as $\mathcal{C}_s(\mathbf{X}, \mathbf{Y})$, is:

**Eq. 36**
$$\mathcal{C}_s(\mathbf{X}, \mathbf{Y}) = \begin{cases} \max_{s_x, s_y} \dfrac{\mathcal{V}_s(\mathbf{X}, \mathbf{Y}; \alpha, s_x, s_y)}{\sqrt{\mathcal{V}_s(\mathbf{X}, \mathbf{X}; \alpha, s_x) \mathcal{V}_s(\mathbf{Y}, \mathbf{Y}; \alpha, s_y)}}, & \text{if } \mathcal{V}_s(\mathbf{X}, \mathbf{X}; \alpha, s_x) \mathcal{V}_s(\mathbf{Y}, \mathbf{Y}; \alpha, s_y) > 0 \\ 0, & \text{if } \mathcal{V}_s(\mathbf{X}, \mathbf{X}; \alpha, s_x) \mathcal{V}_s(\mathbf{Y}, \mathbf{Y}; \alpha, s_y) = 0 \end{cases}$$

In words: for a given exponent parameter $\alpha > 0$, the total similarity coherence is the maximum total coherence with respect to the scale parameters $(s_x, s_y)$. These scale parameter values are used throughout the remainder of this section.

It is of interest to partition the total similarity covariance and coherence into analogues of the real and imaginary parts of the standard complex valued counterparts. To achieve this, we define the real part contribution of the similarity covariance as:

**Eq. 37**  $w_{re} = \mathcal{V}_s(\text{Re}\mathbf{X}, \text{Re}\mathbf{Y}) + \mathcal{V}_s(\text{Im}\mathbf{X}, \text{Im}\mathbf{Y})$

and the imaginary part contribution of the similarity covariance as:

**Eq. 38**  $w_{im} = \mathcal{V}_s(\text{Re}\mathbf{X}, \text{Im}\mathbf{Y}) + \mathcal{V}_s(\text{Im}\mathbf{X}, \text{Re}\mathbf{Y})$

Finally, the total similarity covariance is partitioned into real and imaginary contributions as:

**Eq. 39**
$$\begin{cases} \mathcal{V}_{s,re}(\mathbf{X}, \mathbf{Y}) = \left(\dfrac{w_{re}}{w_{re} + w_{im}}\right) \mathcal{V}_s(\mathbf{X}, \mathbf{Y}) \\ \mathcal{V}_{s,im}(\mathbf{X}, \mathbf{Y}) = \left(\dfrac{w_{im}}{w_{re} + w_{im}}\right) \mathcal{V}_s(\mathbf{X}, \mathbf{Y}) \end{cases}$$

thus giving the additive partitioning:

**Eq. 40**  $\mathcal{V}_s(\mathbf{X}, \mathbf{Y}) = \mathcal{V}_{s,re}(\mathbf{X}, \mathbf{Y}) + \mathcal{V}_{s,im}(\mathbf{X}, \mathbf{Y})$

The total similarity coherence with its additive partitioning into real and imaginary contributions is defined as:

Eq. 41
$$\begin{cases} \mathcal{C}_{s,re}(\mathbf{X},\mathbf{Y}) = \mathcal{V}_{s,re}(\mathbf{X},\mathbf{Y}) / \sqrt{\mathcal{V}_s(\mathbf{X},\mathbf{X})\mathcal{V}_s(\mathbf{Y},\mathbf{Y})} \\ \mathcal{C}_{s,im}(\mathbf{X},\mathbf{Y}) = \mathcal{V}_{s,im}(\mathbf{X},\mathbf{Y}) / \sqrt{\mathcal{V}_s(\mathbf{X},\mathbf{X})\mathcal{V}_s(\mathbf{Y},\mathbf{Y})} \\ \mathcal{C}_s(\mathbf{X},\mathbf{Y}) = \mathcal{C}_{s,re}(\mathbf{X},\mathbf{Y}) + \mathcal{C}_{s,im}(\mathbf{X},\mathbf{Y}) \end{cases}$$

These derivations can be followed in an analogous way in order to produce the "distance coherence" versions.

## 7. Toy examples for real valued data

The functions tested here are the simple straight line, two straight lines forming an "X", the sum of a sine function plus a straight line, random dots on a square, a circle, and a paraboloid. Except for the random dots (where $x$ is random), $x$ is equi-spaced from 0…1. In the case of the two straight lines ("X" shape), the $x$ values are repeated for each straight line. In the case of the paraboloid, $x$ and $y$ are equi-spaced from 0…1. When indicated, noise is added to $y$ only (except for the paraboloid, where noise is added to $z$).

The following figures show the functions:

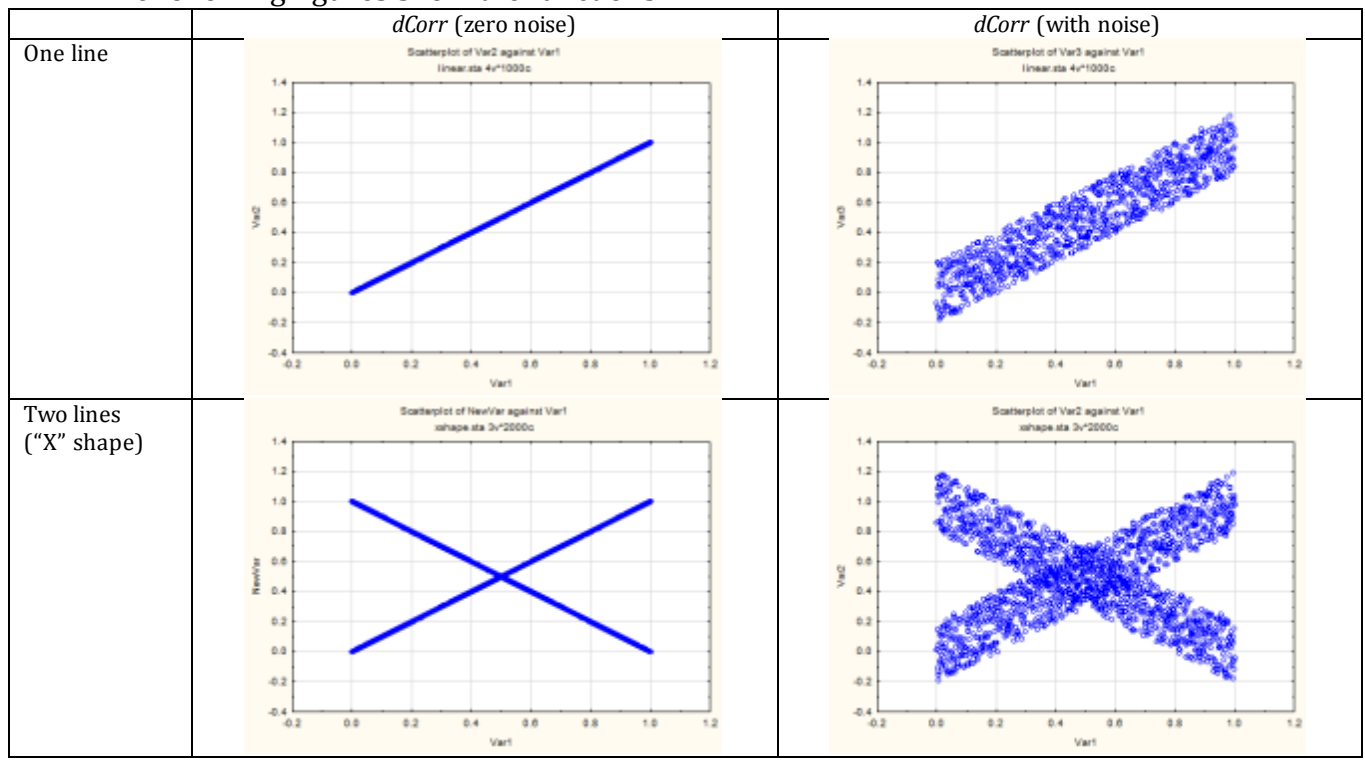

|  | dCorr (zero noise) | dCorr (with noise) |
|---|---|---|
| One line |  |  |
| Two lines ("X" shape) |  |  |
| Line plus sine | 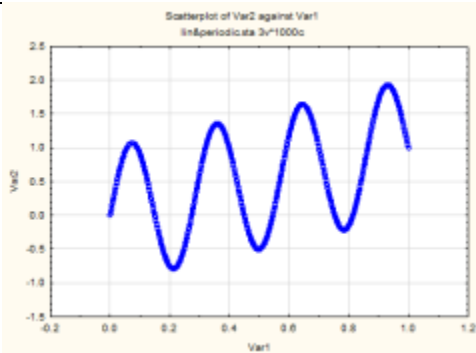 | 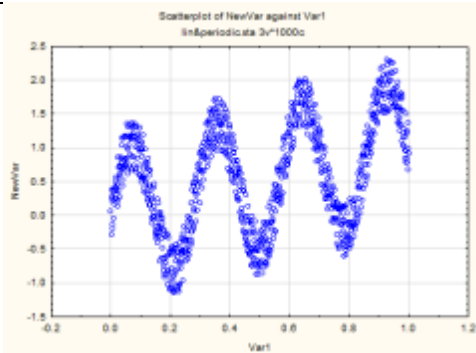 |
| Random dots | 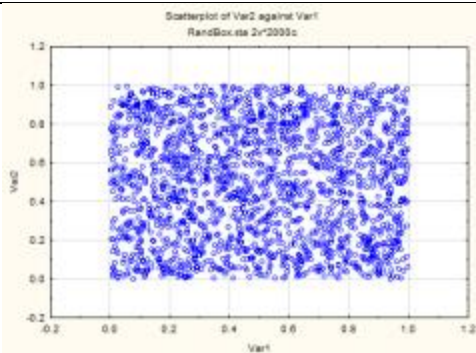 | 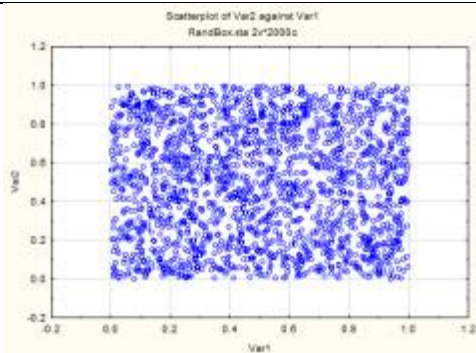 |
| Circle | 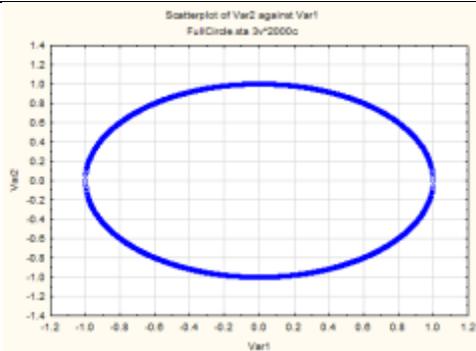 | 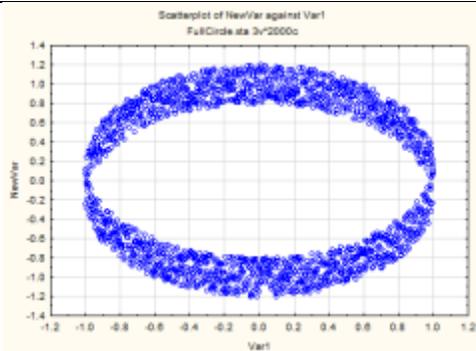 |
| Paraboloid | 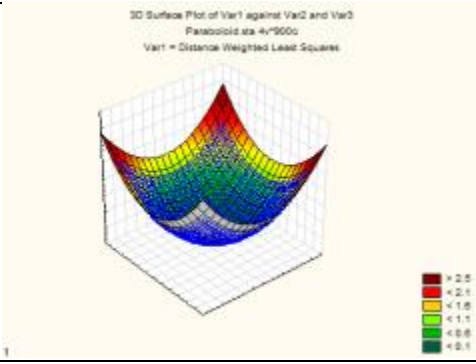 | 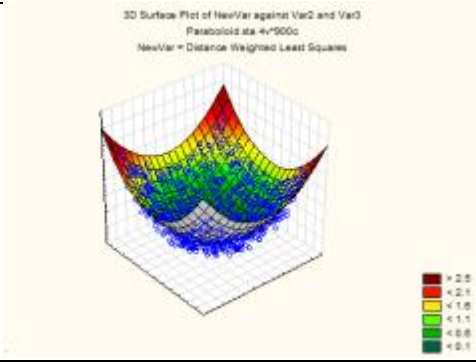 |

### a. Similarity and distance correlation for the toy examples

The following table shows the similarity and distance correlations for the toy examples. In the case of the similarity correlation, the exponent value $(\alpha = 2)$ was used:

| | Functions | Sample size N | $\mathcal{R}_d$ (0 noise) | $\mathcal{R}_d$ ($U[-\beta,+\beta]$ noise $\beta$ in brackets) | $\mathcal{R}_s$ (0 noise) | $\mathcal{R}_s$ ($U[-\beta,+\beta]$ noise $\beta$ in brackets) |
|---|---|---|---|---|---|---|
| One line | $y = x$ | 1000 | 1 | 0.84 [0.2] | 1 | 0.87 [0.2] |
| Two lines ("X" shape) | $y = x$ and $y = 1 - x$ | 2000 | 0.0625 | 0.0372 [0.2] | 0.4728 | 0.1447 [0.2] |
| Line plus sine | $y = \sin(7\pi x) + x$ | 1000 | 0.1584 | 0.1374 [0.4] | 0.4010 | 0.2859 [0.4] |
| Random dots | $x, y \in U[0...1]$ | 2000 | 0.0007 | 0.0007 [0.2] | 0.0031 | 0.0031 [0.2] |
| Circle | $x^2 + y^2 = 1$ | 2000 | 0.0244 | 0.0186 [0.2] | 0.3631 | 0.1327 [0.2] |
| Paraboloid | $z = x^2 + y^2$ | 900 | 0.1234 | 0.0943 [0.4] | 0.2874 | 0.2662 [0.4] |

Note that the two straight lines (with "X" shape), the "line plus sine", and the circle have very low $\mathcal{R}_d$ values. It seems that $\mathcal{R}_d$ favors very much the simple linear relation, and that all non-linear relations will always have very much smaller $\mathcal{R}_d$ values. Although the "X" shape example is not a function, it is the superposition of two functions, and it would be desirable to have a measure that reports a more clearly non-zero value of "association".

In a scale from zero to one, the similarity correlation has a minimal value of 0.29 for the collection of all noiseless toy examples (except the random dots, where no law of association exists between the variables). This minimum value (i.e. 0.29) is clearly distinct from zero, in a scale from zero to one.

In contrast, the distance correlation values are near zero for the two lines ("X" shape) and for the circle. Furthermore, "line plus sine" and paraboloid functions have quite small distance correlation (<0.16).

The following figure shows the similarity correlation as a function of the two scaling parameters (Eq. 18 without taking the maximum), for the "X" shape data. This empirical evidence shows the existence of a single global maximum:

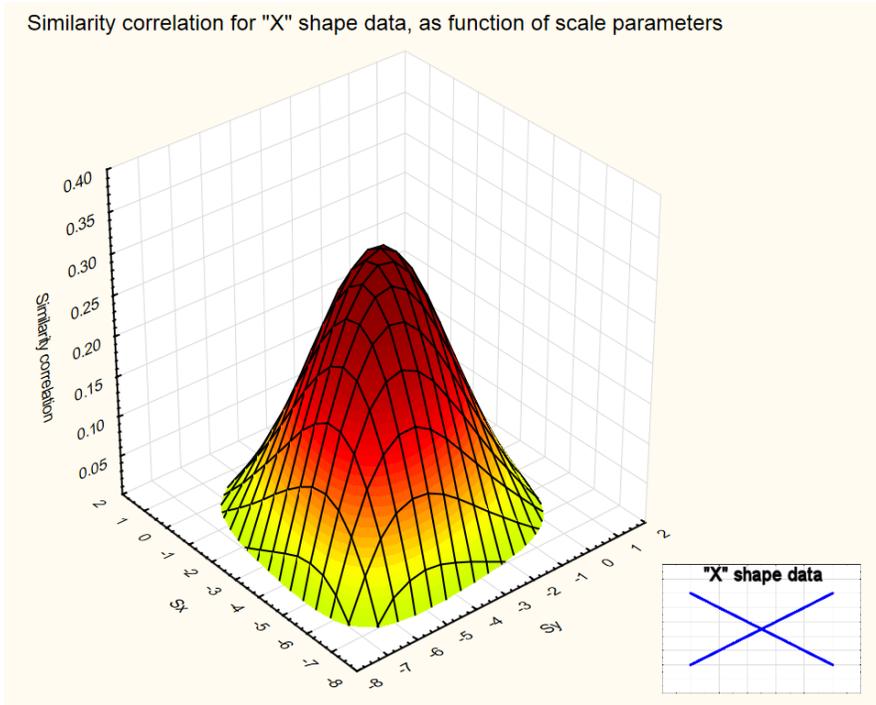

The scale parameters $s_x$ and $s_y$ are shown in logarithmic scale, e.g., a value of -1 on the $s_x$ axis signifies that $s_x = 10^{-1}$.

## 8. Toy examples for complex valued data

In what follows, variables $x$ and $y$, and the coefficient $a$, are complex valued; and the imaginary unit is denoted as $\iota = \sqrt{-1}$.

The functions tested here are:
1. The linear function:

**Eq. 42** $\quad y = ax$

2. The quadratic function:

**Eq. 43** $\quad y = ax^2$

3. The principle square root function:

**Eq. 44** $\quad y = a\sqrt{x}$

where:

**Eq. 45** $\quad \sqrt{x} = \sqrt{\operatorname{Re} x + \iota \operatorname{Im} x} = \sqrt{\dfrac{r + \operatorname{Re} x}{2}} + \iota \operatorname{sgn}(\operatorname{Im} x) \sqrt{\dfrac{r - \operatorname{Re} x}{2}}$

and:

**Eq. 46** $\quad r = \sqrt{(\operatorname{Re} x)^2 + (\operatorname{Im} x)^2}$

4. The principle logarithm:

**Eq. 47** $\quad y = a\ln(x) = a\left[\ln(r) + Arg(x)\right]$

where $Arg(x)$ is required to lie in the interval $[-\pi, +\pi]$.

The following figures display, for each function, the real and imaginary parts of *y* as a function of *x*. In all cases, the coefficient was set to $a = 0.5 - \iota$:

| | $\mathrm{Re}\, y = f(x)$ | $\mathrm{Im}\, y = g(x)$ |
|---|---|---|
| linear $y = ax$ | 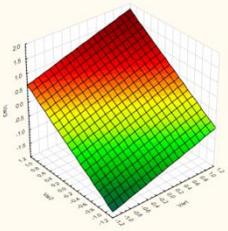 | 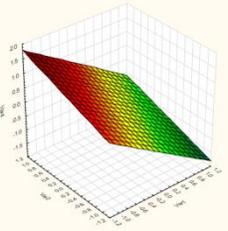 |
| quadratic $y = ax^2$ | 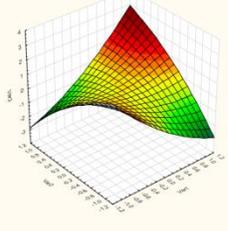 | 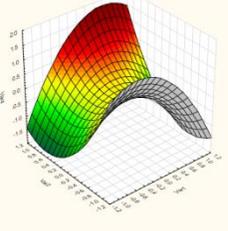 |
| square root $y = a\sqrt{x}$ | 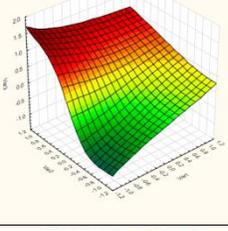 | 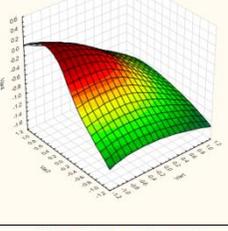 |
| logarithm $y = a\ln(x)$ | 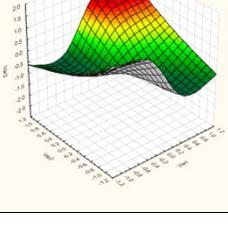 | 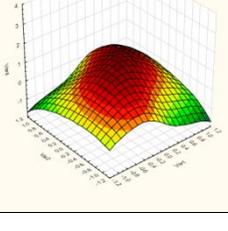 |

### a. Similarity and distance coherence for the toy examples

In a first elementary test, the linear relation of Eq. 42 was used, with two possibilities for the coefficient *a*: pure real or pure imaginary. In such a simple case, the coherences must also be pure real or pure imaginary. Both the similarity and distance coherences satisfied this basic requirement.

Next, for all functions, $\text{Re}(x)$ and $\text{Im}(x)$ were equi-spaced from $[-1...+1]$, on a 30x30 grid, which produced a sample size of 900. The following table shows the results for coherences, where $(\alpha = 2)$ was used for the similarities:

|  | $c_{d,total}$ | $c_{d,re}$ | $c_{d,im}$ | $c_{s,total}$ | $c_{s,re}$ | $c_{s,im}$ |
|---|---|---|---|---|---|---|
| linear $y = ax$ | 1.000 | 0.168 | 0.831 | 1.000 | 0.189 | 0.810 |
| quadratic $y = ax^2$ | 0.202 | 0.101 | 0.101 | 0.563 | 0.281 | 0.281 |
| square root $y = a\sqrt{x}$ | 0.794 | 0.249 | 0.545 | 0.900 | 0.386 | 0.513 |
| logarithm $y = a\ln(x)$ | 0.155 | 0.072 | 0.082 | 0.563 | 0.235 | 0.327 |

In general, the similarity coherences are larger than the distance coherences.

## 9. Applications in functional brain connectivity studies

In general, the non-invasive study of brain function is of great interest. In particular, the study of brain functional connectivity has been an intense area of research in recent years. In many such studies, use is made of measurements of time varying signals of brain activity, and measures of association between the time series are used to assess connectivity. For instance, in fMRI studies, the signals consist of time varying metabolic activity; whereas in studies that use EEG tomographies such as low resolution brain electromagnetic tomography (LORETA) ([8], [9], [10]), the signals consist of time varying electric neuronal activity throughout the cortex.

In the case of high-time resolution stationary signals of electric neuronal activity, it is common to use spectral analysis methods, and to compute coherences between pairs of signals for different frequencies. The new techniques presented in this paper for measures of association between complex valued vectors might prove to be of great value, since they provide a measure of association not limited to linear dependence, and are more general in that the association can be computed for any two groups of multiple signals and not limited only to pairs of single signals.

For instance, consider two distinct brain regions, consisting of *p* and *q* voxels respectively, from which high-time resolution electric neuronal activity signals are available, in the form of *N* non-overlapping time windows. For each time window $i = 1...N$, let $\mathbf{X}_i \in \mathbb{C}^{p \times 1}$ and $\mathbf{Y}_i \in \mathbb{C}^{q \times 1}$ denote the complex valued discrete Fourier transform coefficients at a certain frequency. This data can be used for the computation of the frequency dependent similarity coherence between two brain regions, including its decomposition into real (zero lag) and imaginary (lagged) contributions.

In most previous literature, regions of interest (ROIs) are commonly defined, where the average signal for each ROI is used for computing the (linear) coherence between pairs of ROIs. With the new similarity coherence, there is no need to reduce a ROI to a single signal: all signals within each ROI form one vector, which can then be associated to any other ROI-vector, even of different dimension, and with the further advantage that the association need not be limited to a linear form.

An extreme hypothetical example that illustrates the unique advantages of the similarity coherence consists in calculating a measure of inter-hemispheric association, i.e. to what extent is the activity of the left cortex associated to that of the right cortex. The similarity coherence readily provides such a measure, using information from all voxels from each cortex, without need to form an average signal for each cortex.